\documentclass[conference]{IEEEtran}
\IEEEoverridecommandlockouts
\usepackage{cite}
\usepackage{amsmath,amssymb,amsfonts}
\usepackage{algorithmic}
\usepackage{graphicx}
 \usepackage{lettrine}
\usepackage{textcomp}
\usepackage{xcolor}
\def\BibTeX{{\rm B\kern-.05em{\sc i\kern-.025em b}\kern-.08em
    T\kern-.1667em\lower.7ex\hbox{E}\kern-.125emX}}
\usepackage{amsmath}
\usepackage{amssymb}
\usepackage{amsthm}
\usepackage{amsfonts}
\usepackage{hyphenat}
\usepackage{hyperref}
\usepackage{times}
\usepackage{psfrag}
\usepackage{ifpdf}
\usepackage{array}
\usepackage{multirow}
\usepackage{setspace}
\usepackage{latexsym}
\usepackage{enumerate}
\usepackage{mathtools}
\usepackage{float}
\usepackage{cite}
\usepackage{cleveref}
\usepackage{graphicx}
\usepackage{caption}
\usepackage{epstopdf}
\usepackage{subfig}
\usepackage{accents}
\usepackage{algorithm}
\usepackage{algorithmic}
\usepackage{mathrsfs}
\usepackage{tabularx}
\usepackage{tikz}
\usepackage{pgfplots}
\usepackage{url}  %Required
\usepackage{caption}
\usepackage{rotating}

\definecolor{Fcolor}{rgb}{0,0.5,0.25}

\theoremstyle{definition}

\crefrangelabelformat{equation}{(#3#1#4)\,--\,(#5#2#6)}
\crefmultiformat{equation}{(#2#1#3)}{\,--\,(#2#1#3)}{#2#1#3}{#2#1#3}

\crefname{equation}{}{}
\crefname{figure}{Figure}{Figures}
\crefname{algorithm}{Algorithm}{Algorithms}
\crefname{table}{Table}{Tables}
\crefname{lemma}{Lemma}{Lemmas}
\crefname{theorem}{Theorem}{Theorems}
\crefname{remark}{Remark}{Remarks}
\crefname{section}{Section}{Sections}
\crefname{subsection}{Subsection}{Subsections}
\crefname{definition}{Definition}{Definitions}

\makeatletter
\DeclareRobustCommand{\cev}[1]{%
  \mathpalette\do@cev{#1}%
}
\newcommand{\do@cev}[2]{%
  \fix@cev{#1}{+}%
  \reflectbox{$\m@th#1\vec{\reflectbox{$\fix@cev{#1}{-}\m@th#1#2\fix@cev{#1}{+}$}}$}%
  \fix@cev{#1}{-}%
}
\newcommand{\fix@cev}[2]{%
  \ifx#1\displaystyle
    \mkern#2 1mu
  \else
    \ifx#1\textstyle
      \mkern#2 3mu
    \else
      \ifx#1\scriptstyle
        \mkern#2 2mu
      \else
        \mkern#2 2mu
      \fi
    \fi
  \fi
}

\begin{document}

\title{A Data-Driven Democratized Control Architecture for Regional Transmission Operators\\
%{\footnotesize \textsuperscript{*}Note: Sub-titles are not captured in Xplore andshould not be used}
%\thanks{Identify applicable funding agency here. If none, delete this.}
}
\author{\IEEEauthorblockN{Xiaoyuan Fan}
\IEEEauthorblockA{\textit{Pacific Northwest National Laboratory} \\
Richland, WA USA \\
xiaoyuan.fan@pnnl.gov}
\and
\IEEEauthorblockN{Daniel Moscovitz} % Emanuel typically take one week to get hold of, if not extended, I do not think he will response in time
\IEEEauthorblockA{\textit{PJM Interconnection}\\
Audubon, PA USA\\
daniel.moscovitz@pjm.com}
\and
\IEEEauthorblockN{Liang Du}
\IEEEauthorblockA{\textit{Temple University} \\
Philadelphia, PA USA\\
ldu@temple.edu}
\and
\IEEEauthorblockN{Walid Saad}
\IEEEauthorblockA{\textit{Virginia Tech} \\
Blacksburg, VA USA \\
walids@vt.edu}
}

\maketitle

\begin{abstract}
As probably the most complicated and critical infrastructure system, U.S. power grids become increasingly vulnerable to extreme events such as cyber-attacks and severe weather, as well as higher DER penetrations and growing information mismatch among system operators, utilities (transmission or generation owners), and end-users. This paper proposes a data-driven democratized control architecture considering two democratization pathways to assist transmission system operators, with a targeted use case of developing online proactive islanding strategies. Detailed discussions on load capability profiling at transmission buses and disaggregation of DER generations are provided and illustrated with real-world utility data. By Combining network and operational constraints, transmission system operators can be equipped with new tools built on top of this architecture, to derive accurate, proactive, and strategic islanding decisions to incorporate the wide range of dynamic portfolios and needs when facing extreme events or unseen grid contingencies.
\end{abstract}


\section{Introduction}
The ever-growing penetration of distributed energy resources (DERs), widespread electrification of end-use technologies, and proliferation of digital infrastructure drives a paradigm shift in power system operations, from traditionally being centralized, passive, and rigid to potentially being distributed, active, and flexible. The exogenous dimension of ubiquitous information, such as raw data, local needs, spatial-temporal correlations, and individual conditions, surpass existing decision-making architecture with increasing complexities in multi-scale coordination and control. Such challenges are conveyed to not only distribution system operators but also, and largely, to regional transmission organizations (RTOs). For instance, conventionally centralized, optimization-based, and system-wide voltage regulation schemes for transmission networks have been widely acknowledged as susceptible to single-point failure, communication burden, and scalability issues.

Moreover, U.S. power grids are becoming more vulnerable under increasingly more frequent and extreme weather events. Recently, the U.S. Department of Energy warned that the frequency and intensity of extreme weather hazards may increase, which have already been the largest cause for power outages in U.S. In 2017, major storms cost the U.S. economy an unprecedented amount: over \$200 billion. In 2018, severe weather caused 94 major power outages, 17.5\% and 34.2\% increases compared to 2017 and the average of past 15 years, respectively \cite{OE417}. Recent Texas blackouts have demonstrated the needs (by RTOs) of a prognostic (i.e., proactive and diagnostic) framework that can integrate a multitude of information to proactively diagnose early, detect abnormal asset signatures in electric power grids and dynamically reconfigure assets for  preventive actions.
A vast body of literature \cite{jufri2019state} has been devoted to reactive plans, such as resource allocation and self-healing schemes, to minimize restoration time, reduce financial losses, maximize recovered service area, and enhance grid resiliency. Moreover, despite existing studies on proactive hardening considering weather conditions for transmission networks \cite{bagheri2019resilient} and macro-scale statistical studies using weather data to forecast power outages, such models cannot provide details to RTOs on micro-scale physical assets conditions and end-user needs.
%Current research efforts towards self-healing powers grids aim at fast, economic, and efficient recover from outages. However, T\&D grids are separately analyzed in existing grid hardening and restoration literature, with distribution grids represented as equivalent loads and transmission grids modeled as ideal voltage sources.

Aforementioned literature and efforts did not address one major technical barrier for RTOs to develop prognostic decision-making capabilities, which is the growing information mismatch among system operators, utilities (or transmission owners and generation owners), and end-users. RTOs operate regional markets largely based on historical patterns, forecasted regional demands, and asynchronous information from members in a passive manner. Meanwhile, utilities collect granular data from their assets and co-manage their T\&D networks without knowing the status of generations and individual, active aggregators with diverse economic objectives. As a result, information discrepancy among RTOs, utilities, and prosumers will incite more frequent, unexpected, and susceptible consequences and thus needs to be addressed in a systematic manner.

%Modern outage management systems (OMSs) have been widely deployed in distribution networks to collect customer calls, SCADA data, and advanced metering infrastructure (AMI) readings for situational awareness. Furthermore, fault location, isolation, and service restoration (FLISR) systems are integrated into OMSs to accelerate restoration process \cite{guo2016fault,le2018flisr}. However, most FLISR systems are built on centralized instead of distributed intelligence architectures, which might not operate properly during extreme weather events due to failure of communication or ubiquitous cyber and physical devices.

Furthermore, recent FERC orders 841 and 2222 enable and encourage energy storage systems and DERs to participate in regional organized wholesale, capability, and ancillary markets, and compete with conventional energy resources. However, under extreme conditions and due to lack of system-wide situational awareness, market mechanisms may fail to function. For example, during the 2021 Texas blackouts, ERCOT had experienced widespread loss of generation capability system reserves from not only DERs (e.g., frozen wind turbines or solar panels), but also  thermal units (e.g., lack of fuel or natural gas); system operators eventually had to implement rolling blackouts. With the fast-growing replacement of conventional thermal units by DERs (for instance, there are currently over 10 GW of co-located solar-plus-storage hybrid resources in the PJM capacity market queue \cite{PJM2020solar}), RTOs would further suffer from decreasing situation awareness and grid observability, such as lack of knowledge on actual demand profiles across their service footprints, which motivates this work.

This paper proposes a data-driven democratized control architecture for RTOs to make effective, collective decisions under extreme or unseen conditions. It should be noted that the concept of ``democratized" control has been studied under different contexts but lacks a consistent, unified definition. In \cite{Vidal2018Empirical}, a data-driven empirical study is carried out on 546 air conditioners to evaluate user behaviors and ``democratizes thermostats". Conducted in an unsupervised manner, it is concluded that users can be separated in two groups and there exists a high degree of variability from user to user in their preferences and likely actions. In \cite{zhong2017synchronized}, the democratization of power systems is mostly discussed from the view of how to enable all T\&D stakeholders to play an equal role in power system operations and control, regardless of their sizes or capabilities. However, democratization has not been directly linked to controls. Finally, recent advancements in machine learning, especially federated learning, lead to the concept of \textit{democratized learning} \cite{nguyen2021distributed,nguyen2020self} as a technique to achieve a balance between generalization and specialization in large-scale, distributed machine learning.

To summarize, the above-discussed concepts and frameworks remain inadequate to address the following technical gaps, especially under extreme events or emergent grid conditions.

\begin{itemize}
    \item Spatial Heterogeneity in Demands: Conventional RTO operations are largely based on centralized optimization. Under extreme conditions, centralized dispatch decision are made, typically with only partial system observability, by system operators rather than market operators. In such scenarios, T\&D stakeholders possess inherently variant needs, which are often overlooked. Therefore, the first layer of democratized control is to provide all stakeholder with equal opportunities to be considered, which is different from the equal role as discussed in \cite{zhong2017synchronized}. The proliferation of DER aggregators helps to integrate a large number of spatially dispersed DERs, but the barrier between RTOs and aggregators have yet to be investigated.
    \item Representing the Majority: Under extreme conditions, system operators need to make challenging decisions that benefit the majority, i.e., some stakeholders might take short-term loss. As the second layer of the proposed democratized control, the control objective is not simply benefiting over half of the grid but rather to meet the demands of as many stakeholders as possible. In other words, a 52\% vs 48\% solution by the majority rule (or even the supermajority rule) is not sufficient.
    \item Evolutionary Adaptivity: The third layer of the proposed democratized control is to enable RTOs, T\&D stakeholders, and end-users to evolve and gain enhanced capabilities. On one hand, end-users and prosumers can adaptively reorganize and form different flexibility aggregators. On the other hand, system operators and utilities also need to evolve with the availability of more data and the advancement of computational intelligence as well as adapting to the emerging new norm.
\end{itemize}

\begin{figure}
  \begin{center}
    \includegraphics[width=0.5\textwidth]{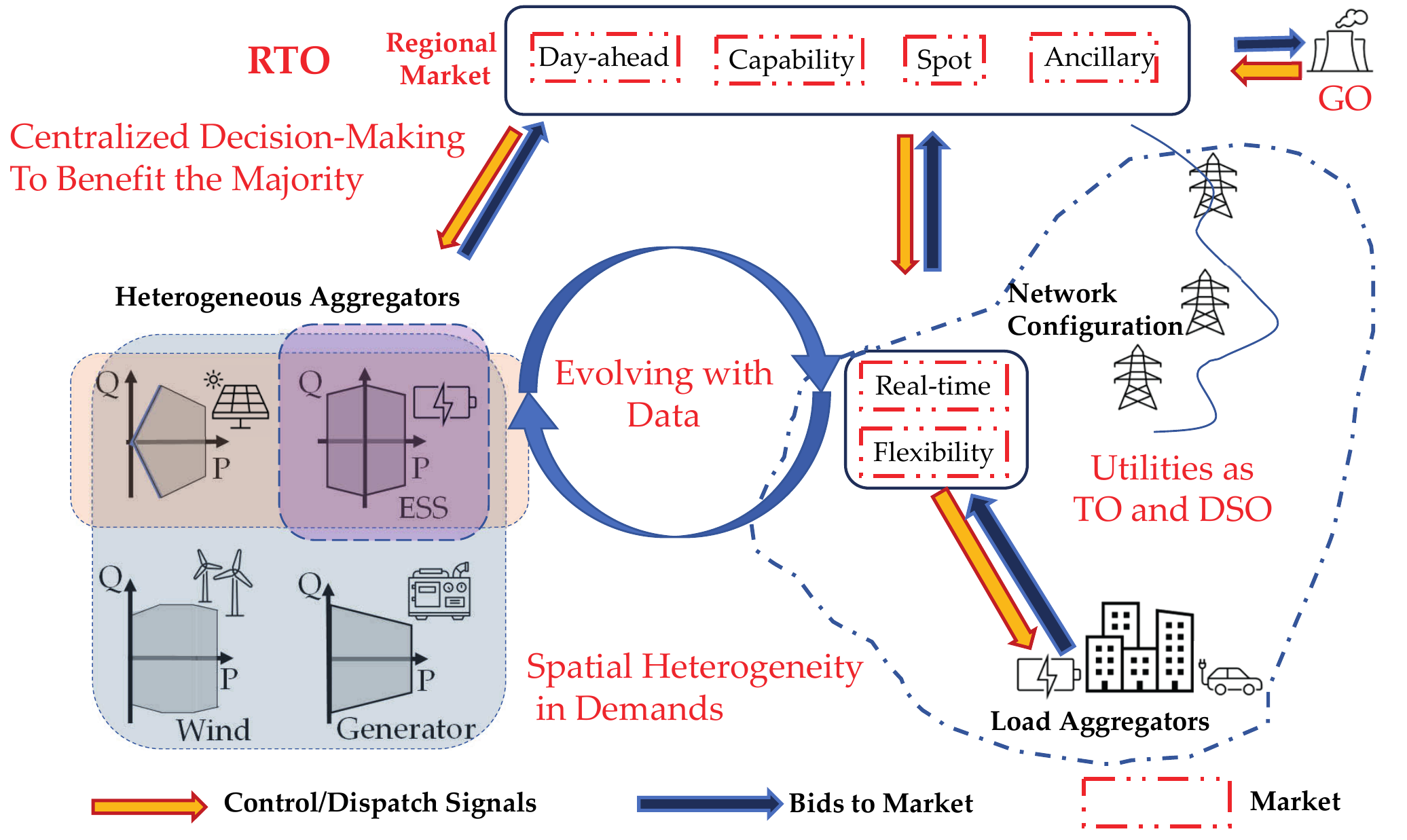}
  \end{center}
  \caption{An illustrative diagram for the proposed architecture.}
    \label{Fig:Overview}
\end{figure}

The hierarchical architecture of the proposed data-driven democratized control is shown in Fig.~\ref{Fig:Overview}. The objective of this paper is mainly on how RTOs should and could carry out democratized decision-makings to allow all stakeholders being represented and considered, and strategically benefit the majority with their best situational estimation based on available data. The proposed data-driven democratized control framework can be implemented in two pathways: 1) without the need of real-time communication, transmission system operators are proposed to carry out non-intrusive load disaggregation techniques to extract and maintain demand profiles and DER generation capabilities at transmission buses, more details are given in Section II; and 2) under extreme events or grid contingencies, with the need of limited communication, system operators are proposed to make centralized, democratized decisions by adopting non-cooperative game-theoretic methods, more details are given in Section III. Section IV presents the strategic islanding usecase, and Section V provides the conclusion and potential future work.

% democratized control is different from democratized learning

% - there could be several groups, majority wins; but minority group might also have a point to make

% - a 52\% vs 48\% win is not a good win, but what can we do?

\section{With Communication: Non-cooperative Game-Theoretic, Centralized Democratization}

To collectively and explicitly incorporate individual stakeholder interests and system operation constraints, two collective decision-making frameworks have drawn researcher's interests in literature: multi-agent systems (MAS) \cite{vrba2014review} and non-cooperative game-theoretic (NGT) methods \cite{saad2012game}. In both frameworks, each agent/player interacts with a defined group of other agents/players and makes independent, autonomous decisions.
%Cooperative control algorithms for MAS typically adopt distributed formulations to achieve collective optimization of global objectives.

Compared to MAS, NGT can incorporate non-smooth or non-convex cost functions \cite{wang2020Noncooperative,PESGMDu2012}, integrate with electricity markets, and achieve emergent global behavior with balanced individual interests in terms of the Nash equilibrium (NE). A NE is a stable state in which all players play the best response to other players' strategic choices and thus no player has a unilateral incentive to deviate. Note that a NE is not necessarily the Pareto optimal and may not even exist. Therefore, one main challenge is to link a NE to optimum of the global objective and converge to this NE, which is known as \textit{NE selection} in games. Researchers have proposed special NGTs, such as the popular \textit{potential games} and \textit{population games} to achieve NE selection capability. One important finding is that cooperative control problems in MAS (e.g., consensus) can be formulated as potential games \cite{monderer1996potential} with guaranteed convergence to a consensus, even with nonconvex constraints \cite{marden2009joint}. In other words, cooperative control in MAS can be considered as a special non-cooperative game-theoretic problem. Another advantage is that NGT methods provide a hierarchical decomposition \cite{li2013designing} between decentralizing global objectives (design of games) and designing local learning algorithms (how to choose or revise strategies).

Under the scope of NGT, T\&D stakeholders are modeled as self-interested decision makers to reflect the fact that they typically represent different owners with diverse economic interests. Specifically, a non-cooperative strategic game consists of
1) A set of players: $\mathcal{P} \coloneqq \{1,...,N\}$;
2) A set of actions for each player $i \in \mathcal{P}$: $\mathcal{A}_i$;
3) An action profile $a \in \mathcal{A}\coloneqq \times_{i \in \mathcal{P}} \mathcal{A}_i$, i.e., the Cartesian product over all players' action sets. Note that $a$ is typically written as $a=(a_i,a_{-i})$, where $a_i$ represents player $\mathcal{P}_i$'s action and $a_{-i}$ denotes all other players' joint action; and
4) Each player $\mathcal{P}_i$ is assigned a payoff function $u_i : \mathcal{A} \to \mathbb{R}$, which in general depends on actions by everyone else.
% For $\mathcal{P}_i$, its best response $BR_i(a_{-i})$ with respect to others' joint actions $a_{-i}$ is defined to be
% \begin{equation}
%     \begin{aligned}
%     BR_i(a_{-i}) =\{ a_i | u_i(a_i,a_{-i}) \geq u_i({a_i}',a_{-i}), \forall {a_i}' \in \mathcal{A}_i\}.
%     \end{aligned}
%     \label{eq:BR_definition}
% \end{equation}
%Note that \textit{nonoperative} indicates that there is no cooperative behavior and thus players compete independently (opposite to group into "coalitions").
Various types of NGT models, such as Stackelberg games and Vickrey-Clarke-Groves (VCG) mechanism designs can be leveraged. The main bottleneck is the lack of guaranteed convergence to a selected NE that is arbitrarily close to the optimizer of system-wide objectives.

Consequently, potential games have gained a lot of interests recently due to their highly structured design to link global objectives and individual interests through the potential function. To be precise, it means that there exists a \textit{potential function} $\phi:\mathcal{A}\to \mathbb{R}$ such that, $\forall i\in\mathcal{I}, \forall a_{-i} \in \mathcal{A}_{-i}, \textnormal{and}~\forall a_i,a_i' \in \mathcal{A}_i$,
\begin{equation*}
    \begin{aligned}
    u_i(a_i,a_{-i}) - u_i(a_i',a_{-i}) = \phi(a_i,a_{-i}) -  \phi(a_i',a_{-i}), \\
    \end{aligned}
\end{equation*}
various distributed learning algorithms in potential games have been applied which guarantee that the player behavior converges to a (possibly suboptimal) NE. However, handling coupled constraints remains a major technical difficulty for applying potential games to power systems.

Compared to the ``winner-take-all" centralized dispatch mechanisms, it has been shown in \cite{du2015game} that the conventional economic dispatch with nonconvex or nonsmooth objections and nonlinear constraints can be solved as an exact potential game, with the capabilities to assign priorities during the solving process. Eventually, the dispatch solutions are presented in the form of NE, which has also been shown to be more fair. For more complicated system-level problems with more contraints, such as the optimized power flow (OPF) problem, state-based potential games have been shown to be feasible to address individual stakeholder needs while achieving the global optimum \cite{liang2018distributed}. Finally, it has been illustrated that all stakeholders can make synchronous, simultaneous decision makings and revisions within potential games \cite{9475962}, which is desired for democratization-type control.

\section{Without Communication: Non-intrusive Load Disaggregation at Transmission Buses}
Conventionally, non-intrusive load disaggregation or non-intrusive load monitoring (NILM) aims at identifying the demand curve of a specific type of loads from the aggregated, measured power consumption curves at the point of measurement. The NILM literature has been largely focused on low-voltage distribution networks or end-use appliances. Recently, the concept of NILM has been extended to disaggregate behind-the-meter (BTM) DER demand
profiles, which are highly correlated with respect to DERs' locations, rate tariffs, population density, and population active time periods \cite{9102286}. For different regions (residential vs commercial, rural vs suburban, and developed vs developing), both demand profiles and BTM DER profiles are fundamentally different. Therefore, the most cost-effective way is to disaggregate the desired BTM DER profile out of the widely available AMI data without extra investment in infrastructures \cite{kabir2019estimation}.

Formally, given an aggregated, measured power consumption profile $x=(x_1,\dots,x_T)$, i.e., a timed sequence of a total of $T$ power consumption data points measured by the AMI, the problem is to determine its corresponding power demand or BTM DER generation profile $P$ (or $P(x)$ if the source power consumption profile $x$ is relevant). Note that the power demand or BTM DER generation profile is considered  aggregated as most smart meters measure the power consumption at the point of measurement  and thus include all loads and BTM generations (i.e., aggregated). Such power demand or BTM DER generation profile is thus a temporal sequence (of the same length $T$) of active power data points. In other words, the value $P_t$ of $P$ at time step $t$ denotes the amount of power demand or BTM DER generation.

Conventionally, RTOs and utilities conduct random sampling and questionnaires to estimate existing BTM solar or solar-plus-storage generation, especially at unregistered locations. With the wide deployment of massive amount of AMIs and the proliferation of advanced data analytic methods using AMI data \cite{wang2018review} and micro-synchrophasors measurement \cite{von2017precision}, several frameworks have been studied to disaggregate BTM solar generations without human-in-the-loop interventions. Change-point detection algorithms to screen out abnormal energy
consumption behaviors from AMI data has been applied to detect unauthorized solar installations and also verified through statistic inferences. Since 2018, many statistical learning methods have been applied to disaggregate BTM solar profiles \cite{chen2017sundance, tabone2018disaggregating,cheung2019towards,vrettos2019estimating} as well as recently to disaggregate BTM solar-plus-storage profiles \cite{cheung2020disaggregation}. A common drawback is their complexity to train and learn, the lack of generalization capability, and computational burden at the device level.

This paper proposes to extend the concept of non-intrusive disaggregation to utilize available demand profiles and DER generation data at RTOs and enable RTOs with data-driven capabilities to estimate and maintain their granular understanding of demand profiles and DER generation profiles (such as the MW-scale solar-plus-storage, also called hybrid DC resources), without explicitly and constantly require data exchange between RTOs and market participants or T\&D stakeholders. In other words, non-intrusive load disaggregation techniques, which have been proven to be applicable and feasible in distribution networks, can be naturally extended to establish demand and DER generation profiles at transmission buses in a timely manner, no extra metering infrastructure is required.

\begin{figure}
  \begin{center}
    \includegraphics[width=0.5\textwidth]{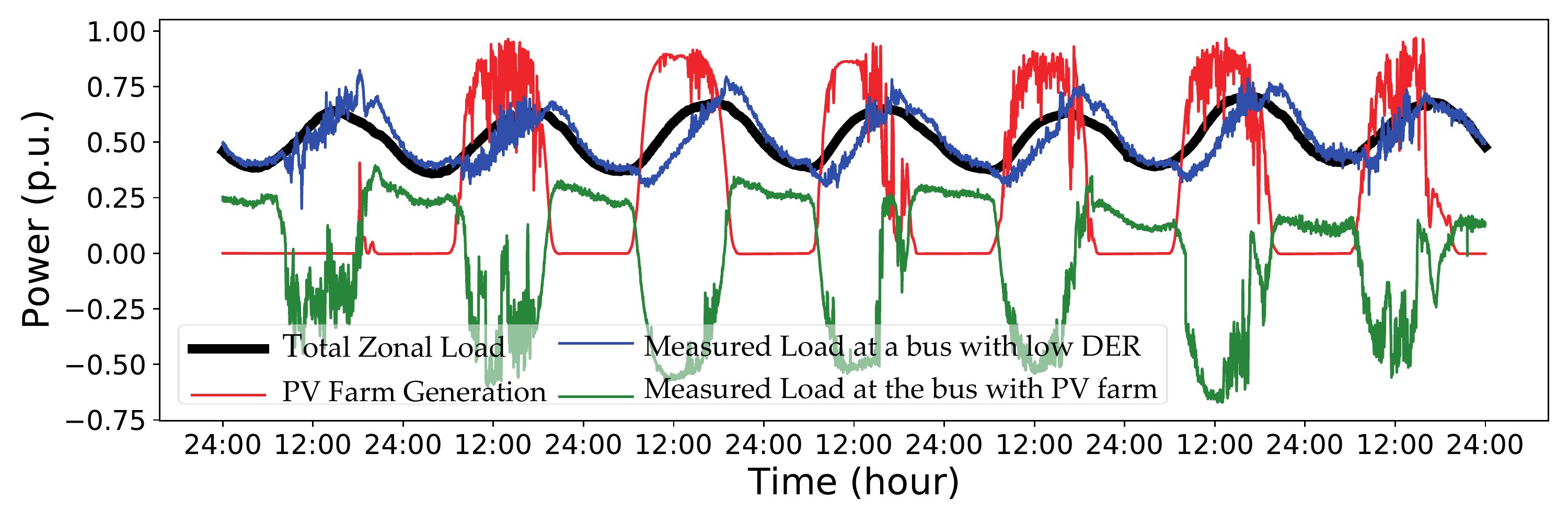}
  \end{center}
  \caption{Illustration of real-world pre-processed, minute-level, normalized, and anonymized utility data.}
    \label{PJM_data}
\end{figure}

Fig.~\ref{PJM_data} illustrates real-world pre-processed, normalized, and anonymized load profiles at several transmission buses with solar farm generation data available to RTOs. A sample plot of weekly data between July 1 and July 7, 2020, presents the total zonal demand of an anonymized utility zone (in black), a bus with low solar penetration (in blue), and another bus with dominating solar generations (in red) with its actual demand curve (in green). It can be observed that load patterns at buses with low solar generation closely follow the zone, while load patterns high solar penetration are often dominated by it and thus recorded as volatile negative injections, i.e., not explicitly presenting useful information by themselves.

\begin{figure}
  \begin{center}
    \includegraphics[width=0.45\textwidth]{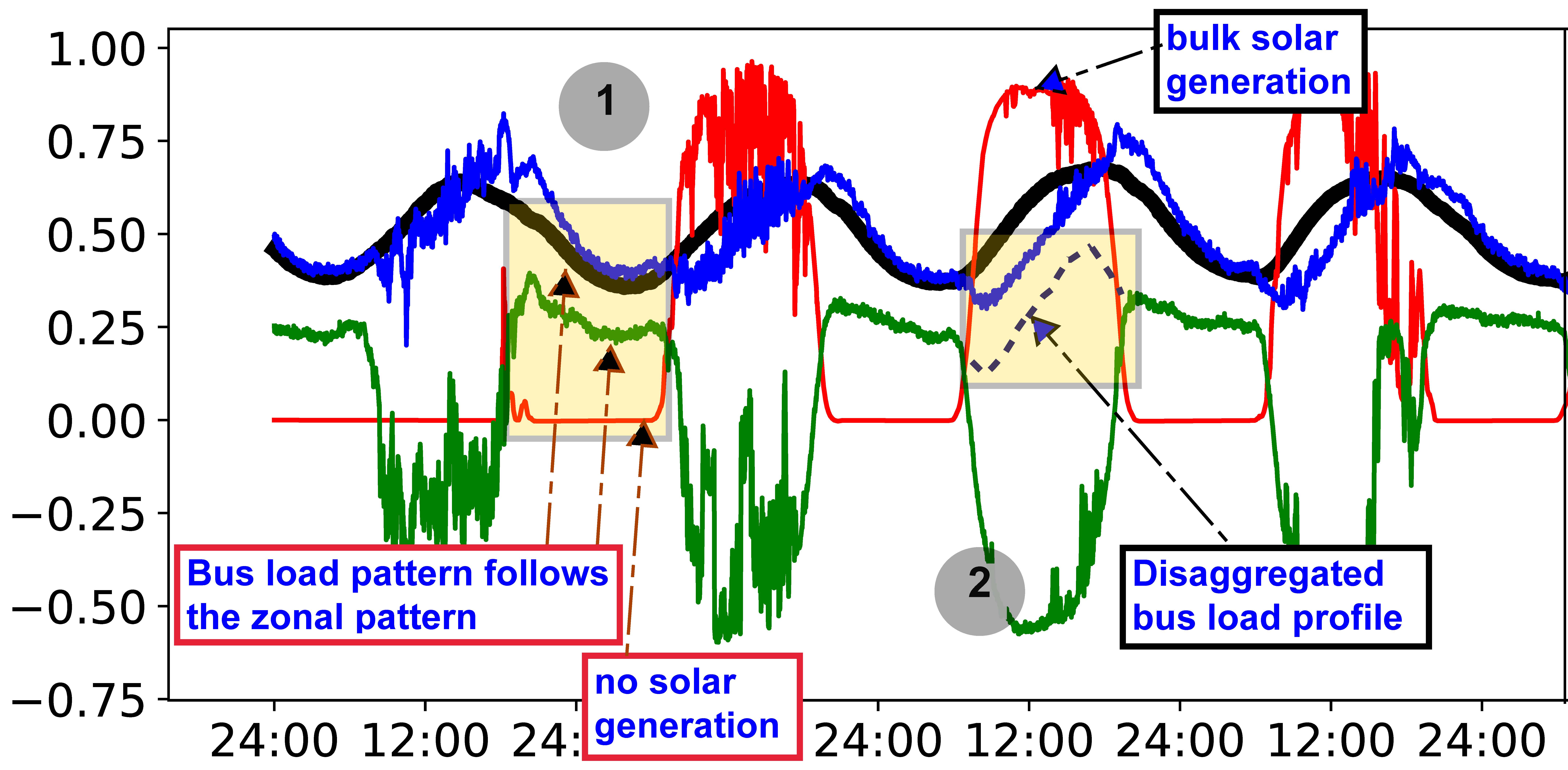}
  \end{center}
  \caption{Non-intrusive data-driven disaggregation of load profiles with two scenarios illustrated.}
    \label{Fig:bus_load_disaggregation}
\end{figure}

To illustrate the feasibility and applicability of data-driven disaggregation of (measured, aggregated) load profiles in a non-intrusive manner to assist system operators, Fig.~\ref{Fig:bus_load_disaggregation} (based on RTO data plotted in Fig.~\ref{PJM_data}) provides a concrete example. The following
two scenarios on the measured load profile (i.e., aggregated signals of local demands and generations, shown by the green line) of an anonymized, real-world bus with significant solar generation capability can be observed and compared:

\begin{itemize}
    \item Scenario 1: With no solar generation (e.g., at night time, and thus the red line remains zero), it can be observed that the measured load profile presents as demands and closely follows the zonal load profile (the black line) and the other bus (the blue line);
\item Scenario 2: With significant amount of solar generation at daytime, the measured load profile is dominated by solar generation and thus presents as volatile injections (i.e., negative demands). Consequently, the actual demands at this bus during daytime remains unknown, without extra investment in sensing infrastructures across the grid.
\end{itemize}

The expected outcome is also illustrated in Scenario 2 of Fig.~\ref{Fig:bus_load_disaggregation}, in which the actual demand is dominated by the solar generation (and thus unavailable) and the artificially estimated demand (shown by the dashed line) is expected to stochastically aligned with the zonal demand patterns as well as historical demand profiles from other days.

Under extreme events or grid contingencies, DER generation may deviate from normal patterns in an uncontrollable manner or even fail (e.g., the lost DER generation capability from frozen wind turbines and solar panels during 2021 Texas blackouts). With the fast-growing replacement of conventional thermal units by DERs, RTOs would further suffer from decreasing situation awareness and grid observability, such as lack of knowledge on off-the-chart demand profiles across their service footprints, as well as irregular time shift due to unforeseeable change of residential, commercial and industry customer behaviors.
% \begin{figure}
%   \begin{center}
%     \includegraphics[width=0.5\textwidth]{figures/Bus_Load_Disaggregation.jpg}
%   \end{center}
%   \caption{Non-intrusive data-driven disaggregation of load profiles with two scenarios illustrated.}
%     \label{Fig:bus_load_disaggregation}
% \end{figure}

\section{Illustrating Example: Strategic Islanding}
In response to various grid emergencies due to various external disturbances and equipment failure,  grid operators heavily rely on off-line model-based reliability study \cite{BS2020Islanding, XFEIOscillation2021} and pre-configured system protection schemes \cite{XF2011Event2020} to minimize the interruption of energy services. But the growing challenges imposed by natural disasters call upon the whole power industry and research community to evaluate the shortcoming of existing methodologies. For example, conventional reactive islanding mechanisms based on operation manuals and reliability-driven protection equipment have been challenged when combating extreme events and alleviating recent rolling blackouts in Texas and California. During such extreme events, system operators, as the islanding decision-makers, often lack grid awareness and observability of the heterogeneous and diverse flexibilities, capabilities, and needs over their vast expanse of geographic footprints. As a result, to combat natural disasters and alleviate more frequent, unexpected, and susceptible consequences due to those extreme operation conditions, operators need to be equipped with new tools and analytics to derive accurate, proactive, and strategic islanding decisions to incorporate the wide range of dynamic portfolios and needs when facing extreme events.

The proposed data-driven democratized control architecture fulfills such gap, by incorporating evolutionary load capability profiling and democratized control, to assist grid operators in response to grid emergencies while maximizing the benefits of high DER penetration and empowered prosumers. By bridging the three gaps listed in Section I, which include 1) spatial heterogeneity in Demands, 2) representing the majority, 3) evolutionary adaptivity, the decision-making process of grid operators for proactive islanding can be boosted with two democratization pathways, namely non-cooperative centralized democratization and non-intrusive load disaggregation, as well as the  data-driven architecture as a whole.

More specifically, real-world utility data, e.g., interconnection models \cite{BS2020Islanding, XFEIOscillation2021,XF2011Event2020}, zonal and individual buses’ demand profiles shown in Fig.~\ref{PJM_data}, and major DER generation profiles shwon in Fig.~\ref{Fig:bus_load_disaggregation}, can be leveraged by the state-of-art graph learning and reinforcement learning techniques and inform the collective profiling of interests and capabilities across an RTO's service footprint; various entities' economic preferences and societal objectives could also be inclusively considered. Such online proactive islanding analysis not only can provide grid operators with high-confidence grid portrait based on  representative temporal and spatial patterns, but also include and reflect the democratized control participants' willingness and capability into the strategic islanding decisions.

\section{Conclusions and Future Work}
This paper proposed a data-driven democratized control architecture with two democratization pathways to assist system operators, and further applied this architecture to develop one use case considering online proactive islanding strategies. Real-world utility data are included for detailed illustration of load capability profiling at transmission buses and disaggregation of DER generations.
%with a targeted use case of developing online proactive islanding strategies. Detailed discussions on load capability profiling at transmission buses and disaggregation of DER generations were provided and illustrated with real-world utility data. By Combining network and operational constraints, operators can be equipped with new tools built on top of this architecture, to derive accurate, proactive, and strategic islanding decisions to incorporate the wide range of dynamic portfolios and needs when facing extreme events.
Future work includes validation of the proposed architecture with real-world utility model and data, performance evaluation and comparison with existing solutions, and industry engagements with utility and grid software vendors.
\bibliographystyle{IEEEtran}
\bibliography{reference}
\end{document}